\documentclass[ aps,%
floatfix,%
final,%
notitlepage,%
oneside,%
onecolumn,%
nobibnotes,%
nofootinbib,%
superscriptaddress,%
showpacs,%
showkeys
]%
{revtex4}

\usepackage{graphicx}

\begin{document}

\title{Measurement of the $\eta - \eta'$ mixing angle in $\pi^{-}$ and $K^{-}$ beams with GAMS-$4\pi$ Setup}

\author{S.V.~Donskov}
\author{V.N.~Kolosov}
\author{A.A.~Lednev}
\author{Yu.V.~Mikhailov}
\author{V.A.~Polyakov}
\author{V.D.~Samoylenko}
\email{Vladimir.Samoylenko@ihep.ru}
\author{G.V.~Khaustov}
\affiliation{Institute for High Energy Physics, Protvino, Russia}

\begin{abstract} 
The results of mixing angle measurement for $\eta'$, $\eta$ mesons generated in charge exchange reactions 
with $\pi^{-}$ and  $K^{-}$ beams are preseneted.
When the $\eta'$, $\eta$ mesons are described in nonstrange(NS)--strange(S) quark basis the
$\pi^{-}$ and  $K^{-}$ beams allow to study $|\eta_{q}\rangle$ and  $|\eta_{s}\rangle$ parts of
the meson wave function. 
The cross section ratio at $t'=0$ (GeV/c)$^{2}$ in the $\pi^{-}$ beam is
$R_{\pi}(\eta'/\eta)= 0.56 \pm 0.04$ and result in mixing angle $\phi_{P} = (36.8 \pm 1.)^{o} $.
For $K^{-}$ beam the ratio is $R_{K}(\eta'/\eta)= 1.30 \pm 0.16$.
It was found that gluonium content in $\eta'$ is $\sin^{2}\psi_{G}= 0.15 \pm 0.06$.
The experiment was carried out with GAMS-4$\pi$ Setup.
\end{abstract}
\pacs{
13.25.Jx,	
13.75.-n,
13.75.Lb
}
\keywords{mixing angle, pseudoscalar, gluonium}
\maketitle 
\section{Introduction}

 The investigation of the meson structure started during the second part of the last century, when
in the  native quark model  frame the experimental consequence of the $SU(3)$
broken symmetry had been measured. In this model the physical states
$\eta$ и $\eta'$ are the linear combination of  $SU(3)$ singlet and nonet:
\begin{equation}
\left(
\begin{array}{c}
\eta \\
\eta'
\end{array}
\right)
 = 
\left(
\begin{array}{cc}
\cos\theta_{P} &  -\sin\theta_{P} \\
\sin\theta_{P}  & \cos\theta_{P} 
\end{array}
\right)
\left(
\begin{array}{c}
\eta_{8} \\
\eta_{0}
\end{array}
\right)
\label{mix18}
\end{equation}
where
\begin{equation}
|\eta_{0} \rangle = \frac{1}{\sqrt{3}} |u\bar{u} + d\bar{d} + s\bar{s} \rangle,\mbox{ }
|\eta_{8} \rangle = \frac{1}{\sqrt{3}} |u\bar{u} + d\bar{d} - 2s\bar{s} \rangle,
\end{equation} 
and  $\theta_{P}$ --- the mixing angle in singlet-octet representation.
In nonstrange(NS)--strange(S)  quark base
\begin{equation}
 |\eta_{q}\rangle =\frac{|u\bar{u} + d \bar{d}\rangle}{\sqrt{2}},\mbox{ }
 |\eta_{s}\rangle = |s\bar{s}\rangle 
\label{qbase}
\end{equation}
the mixing becomes
\begin{equation}
\left(
\begin{array}{c}
\eta \\
\eta'
\end{array}
\right)
 = 
\left(
\begin{array}{cc}
\cos\phi_{P} &  -\sin\phi_{P} \\
\sin\phi_{P}  & \cos\phi_{P} 
\end{array}
\right)
\left(
\begin{array}{c}
\eta_{q} \\
\eta_{s}
\end{array}
\right) ,
\label{mixq}
\end{equation}
and the angles $\theta_{P}$  and  $\phi_{P}$
are related through
\begin{equation}
\theta_{P} = \phi_{P} - \arctan{\sqrt{2}}.
\label{anglelink}
\end{equation}

But it is appear that the internal structure of the pseudoscalars
is not  so simple, and the outlined approach is simplified \cite{feldmann}. 
The experimental properties of the $\eta'$ can be explained by  assuming
that there is a gluonium component 
in the meson wave function \cite{vainshtein, kataev1, kataev2, likhoded}. 
KLOE used \cite{kloe} two angles method to describe  $\eta$, $\eta'$ wave functions: 
\begin{equation}
\begin{array}{llll}
\eta' = & \cos\psi_{G} \sin\psi_{P}|\eta_{q}\rangle & + \cos\psi_{G} \cos\psi_{P}|\eta_{s}\rangle & +\sin\psi_{G}|G\rangle \\
\eta  = & \cos\psi_{P}|\eta_{q}\rangle              & - \sin\psi_{P} |\eta_{s}\rangle,             &  
\end{array}
\label{withgluon}
\end{equation} 
where $\psi_{P}$ is the $\eta - \eta'$ mixing anlge and $\sin^{2} \psi_{G}$ is the 
gluonium fraction in $\eta'$. The KLOE result is $\sin^{2}\psi_{G}= 0.12 \pm 0.04$. 
\begin{figure}[htb]
\includegraphics[width=0.8\textwidth]{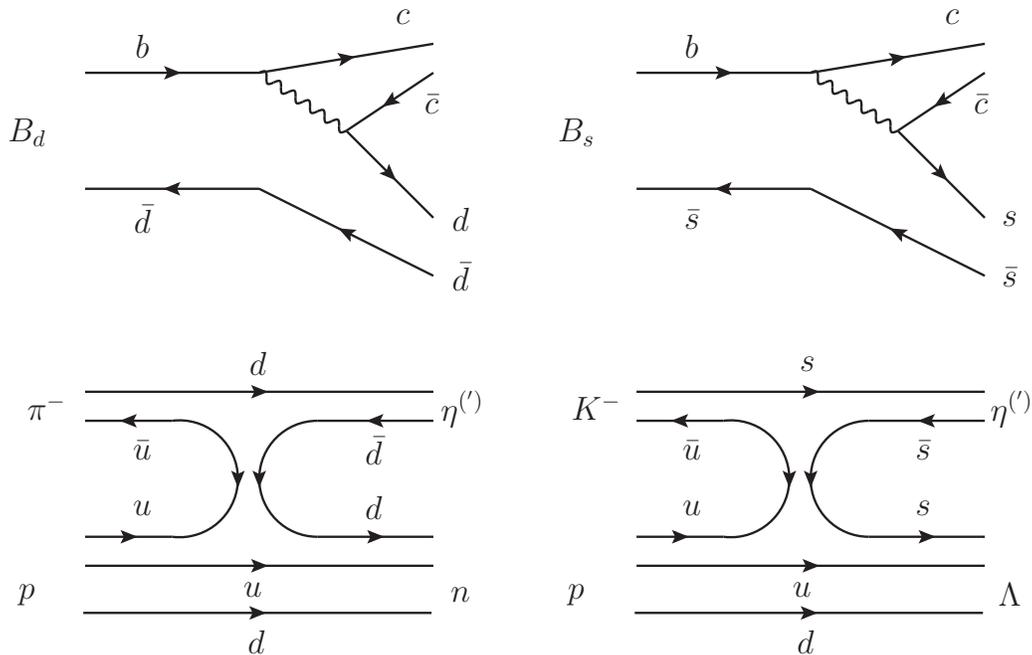}
\label{fig_diag}
\caption{The diagramms of the decay $B_{d}, B_{s}$ mesons (top)
and charge-exchange reaction in  $\pi^{-}$ and   $K^{-}$ beams (bottom).}
\end{figure}

As it was being understand in the  process of the study of the meson structure,
the accuracy of the mixing angle measurement can be improved if 
to prepare the known initial  quark states ( $|\eta_{q}\rangle$ or $|\eta_{s}\rangle$ ),
as in the $B_{s}, B_{d}$ meson decays~\cite{bmesondecay}, Fig.~\ref{fig_diag}a.
The production $\eta$, $\eta'$ meson in the charge-exchange reaction with  $\pi^{-}$
and $K^{-}$ beam can be represented by planar diagrams, Fig.~\ref{fig_diag}b.
These diagramms are dominated, and  the initial state 
$d\bar{d}$ ( $|\eta_{q}\rangle$ ) shall have been  prepared in $\pi^{-}$ beam 
and $s\bar{s}$ ( $|\eta_{s}\rangle$ ) state in the  $K^{-}$ beam.
In this approach the ratio of the $\eta'$, $\eta$ cross sections for $\pi^{-}$ beam is
\begin{equation}
R_{\pi} =  \frac{\sigma(\eta')}{\sigma(\eta)}  = \tan^{2}\phi_{P} 
\label{Rpi}
\end{equation}
and
\begin{equation}
R_{K} =  \frac{\sigma(\eta')}{\sigma(\eta)} = \cot^{2}\phi_{P}
\label{RK}
\end{equation}
for $K^{-}$ beam. The idea  of measuring the mixing angle by cross sections ratio  was proposed
for the first time in  \cite{lipkin}.  
Experimental results obtained in  $\pi^{-}$ beam were published in \cite{edwards, nice_etap}.
To remove the mixing angle dependencies from kinematic factors and phase space
the measurement is performed at $t'= t - t_{min}=0$ 
\begin{equation}
R =   \frac{d\sigma(\eta')}{dt}/\frac{d\sigma(\eta)}{dt}   \mid_{t=0}.
\label{Rdef}
\end{equation}
If the differential cross section parameters are the same,  then $R$
equal  (\ref{Rpi}) or (\ref{RK}) depending from the beam particle type.

The results of the mixing angle measurements in $\pi^{-}$ and $K^{-}$ beam 
for pseudoscalar mesons $\eta$, $\eta'$ are presented in our report. 
The dependencies   $R_{\pi}, R_{K}$ vs $t$ are shown also.
The data analysis had been performed as in simple quark model and with gluon content.
\section{GAMS-$4\pi$ Setup and Event selection}

The main detector of GAMS-$4\pi$ Setup is the lead glass electromagnetic calorimeter
\cite{gams}.
The central part of the calorimeter contains  PWO crystalls in order 
to improve energetic and spatial resolution \cite{pwo}. 
The detailed description of the experiment performance and data processing 
has been given elsewhere  \cite{gams4pi}.
The charged particle beam of IHEP U70 accelerator consists
of 98\%  $\pi^{-}$ and 2\% $K^{-}$ mesons, the admixture of other particles is negligible.
The type of the beam particle is defined reliable by two threshold cherenkov 
counters \cite{pps} with quartz optics.
The charge-exchange reaction at a 32.5 GeV/c   beam momentum was
used as the source of the monoenergetic pseudoscalar mesons
\begin{equation}
\pi^{-} (K^{-})\mbox{ } p \rightarrow M^{0} \mbox{ }n (\Lambda)
\label{cex}
\end{equation}
the $M^{0}$ state decaying into photons. The study of the mesons production
was performed in $2\gamma$  decay mode, that significally reduced the systematic
errose at the ratio cross sections measurements.
\begin{figure}[htb]
\includegraphics[width=0.90\textwidth]{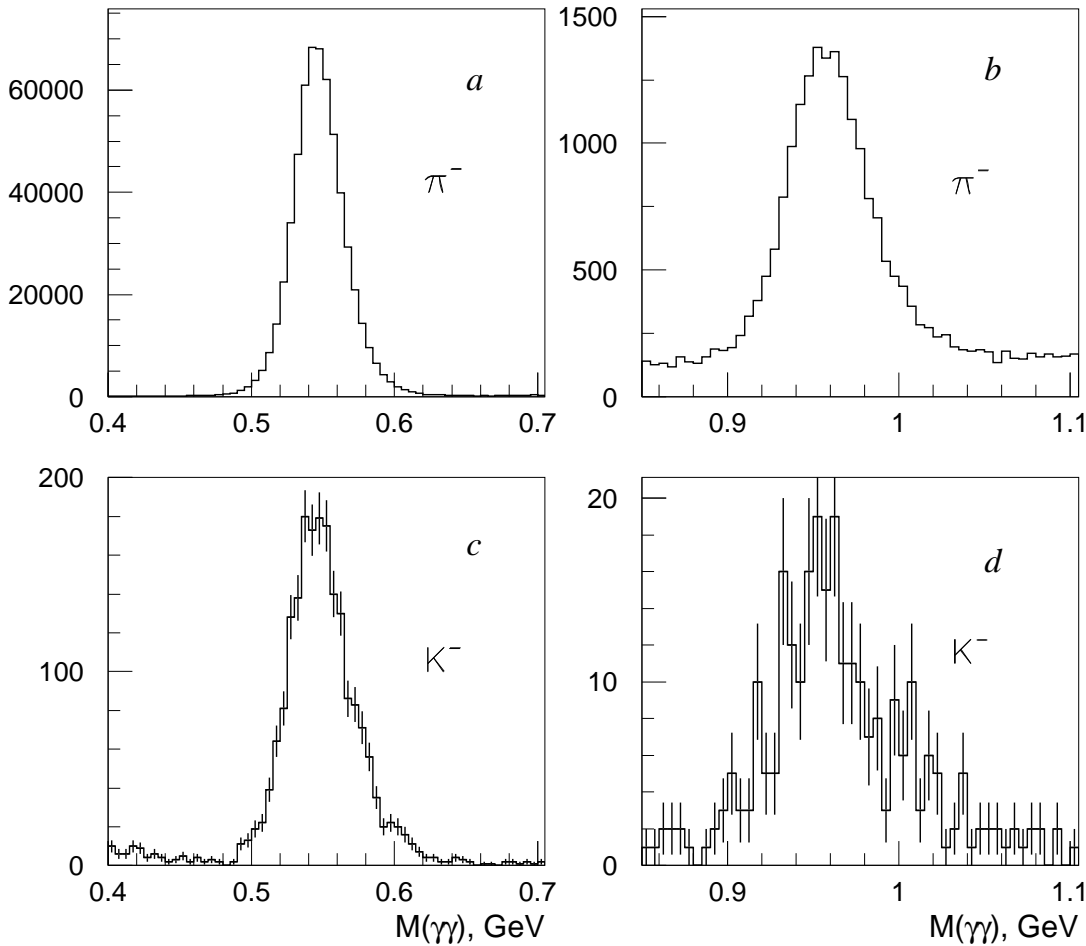}
\caption{Experimental  $2\gamma$ events mass spectrum  in $\eta$ and $\eta'$ regions
for $\pi^{-}$ beam ( histos {\it a} and  {\it b}  accordingly) 
and for $K^{-}$ beam ( {\it c} and  {\it d} ).}
\label{fig2g_pik}
\end{figure}
The experimental mass spectrum  are shown in Fig.~\ref{fig2g_pik}.
The $0.66 \cdot 10^{6}$ events of the decay $\eta \rightarrow 2 \gamma$ 
and $ 14.27 \cdot 10^{3}$ of $\eta' \rightarrow 2 \gamma$
in $\pi^{-}$ beam were detected. For $K^{-}$ beam the number of events
are $2.22 \cdot 10^{3}$ and $0.21 \cdot 10^{3}$ accordingly.
A 2C fit (the additional constrain 
$M_{\gamma \gamma} = M_{\eta}$  or $M_{\gamma \gamma} = M_{\eta'}$)
 was performed for t-dependence study  to improve the resolution.
\section{Mixing angle measurement}

\begin{figure}[htb]
\includegraphics[width=0.90\textwidth]{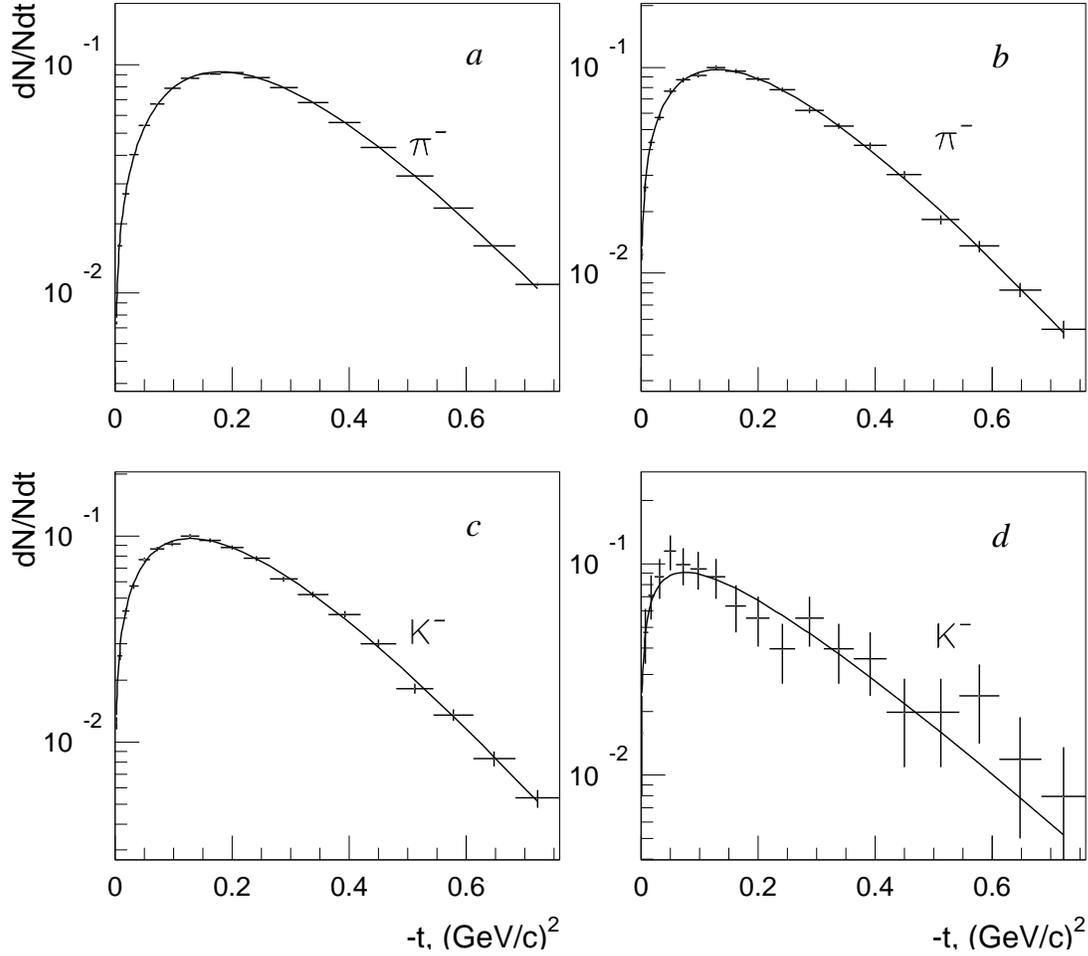}
\caption{Normalized differential cross section $\eta$ and $\eta'$
mesons with fitted function (\ref{dsdt})
in the $\pi^{-}$ beam ( {\it a} and {b}) and in the $K^{-}$ beam ( {\it c} and {\it d} ).}
\label{fig2g_t}
\end{figure}

A phenomenological function, according to \cite{nice_eta}, is used to describe
the differential cross section 
\begin{equation}
\frac{d\sigma}{dt} =\frac{d\sigma}{dt}\mid_{t=0} (1 - gct) e^{ct},
\label{dsdt}
\end{equation} 
where $g= \sigma_{-}/\sigma_{+}$  is the ratio flip and non-flip cross section.
The fit results are shown in Tabl.~\ref{Tablefit} and in Fig.~\ref{fig2g_t}.
Its interesting  to see that in $K^{-}$ beam the non-flip amplitude dominates.
%
\begin{table}
\caption{The results of differential cross section  fit by the function
(\ref{dsdt}) in $\pi^{-}$ and  $K^{-}$ beams
($a \equiv \frac{d\sigma}{dt}\mid_{t=0}$).}
\begin{center}
\begin{tabular}{|c|ccc|ccc|} \hline
\multicolumn{4}{|c|}{$\eta$} & \multicolumn{3}{c|}{$\eta'$} \\  \hline
          & $a$              & $c$             & $g$            &  $a$            &     $c$      &         $g$   \\
$\pi^{-}$ & $3.61 \pm 0.04 $ & $7.9 \pm 0.1 $ & $2.5 \pm 0.1$  & $4.58 \pm 0.14$ & $8.9 \pm 0.2$& $1.9 \pm 0.2$ \\ 
$K^{-}$   & $14.2 \pm 0.7 $ & $9.8 \pm 1.2$   & $0.15 \pm 0.17$& $8.3  \pm 0.9$  & $6.1 \pm 2.1$& $0.0 \pm 0.4$ \\
\hline
\end{tabular}
\end{center}
\label{Tablefit}
\end{table}
%

The cross section ratio in $\pi^{-}$ beam at $t'=0$ is
\begin{equation}
R_{\pi} =   \frac{d\sigma(\eta')}{dt}/\frac{d\sigma(\eta)}{dt}  \mid_{t=0} = 0.56 \pm 0.04,
\label{R_pi_t0}
\end{equation}
and the errore is due to mainly by the accuracy of the effiency calculation at small $t$.
The  obtained result is in agree with NICE experiment $R_{\pi}(\eta'/\eta) = 0.55 \pm 0.06$ \cite{nice_etap}).
The mixing anlge in the quark basis is
\begin{equation}
\phi_{P} = (36.8 \pm 1.0)^{o},
\label{theta_pi_t0}
\end{equation}
and corresponds  $\theta_{P} = -17.9 \pm 1.0$  in the singlet-octet presentation.

For $K^{-}$ beam the ratio is
\begin{equation}
R_{K} =   \frac{d\sigma(\eta')}{dt} /\frac{d\sigma(\eta)}{dt}  \mid_{t=0}  =  1.30 \pm 0.16.
\label{figR_K_t0}
\end{equation}

From the native quark model we can expect
\begin{equation}
R_{\pi} \cdot R_{K} = \tan^{2}{\phi_{P}} \cdot \cot{\phi_{P}}^{2} = 1,
\label{Rmult}
\end{equation}
but the experimental data lead to
\begin{equation}
R_{\pi} \cdot R_{K} = 0.72 \pm 0.09.
\label{multp}
\end{equation}
This discrepance can be  explained  if the internal structure of the pseudoscalars contains 
a some part, which did not consider in the model, or if the planar diagram has limited
application for the investigated processes.
If we follow the KLOE approach (\ref{withgluon}) then the next relations are obvious
\begin{equation}
\begin{array}{c}
R_{\pi} = \cos^{2}{\psi_{G}} \tan^{2}{\phi_{P}} \\
R_{K} = \cos^{2}{\psi_{G}} \cot^{2}{\phi_{P}},
\label{kloe_R}
\end{array}
\end{equation}
and
\begin{equation}
R_{\pi} \cdot  R_{K} = \cos^{4}{\psi_{G}}.
\label{RR}
\end{equation}
The gluonium content of the  $\eta'$ mesons from our data is
\begin{equation}
\sin^{2}{\psi_{G}} = 0.15 \pm 0.06,
\label{gluonp}
\end{equation}
The KLOE result based on radiative meson decays is $0.12 \pm 0.04$.

\begin{figure}[htb]
\includegraphics[width=0.90\textwidth]{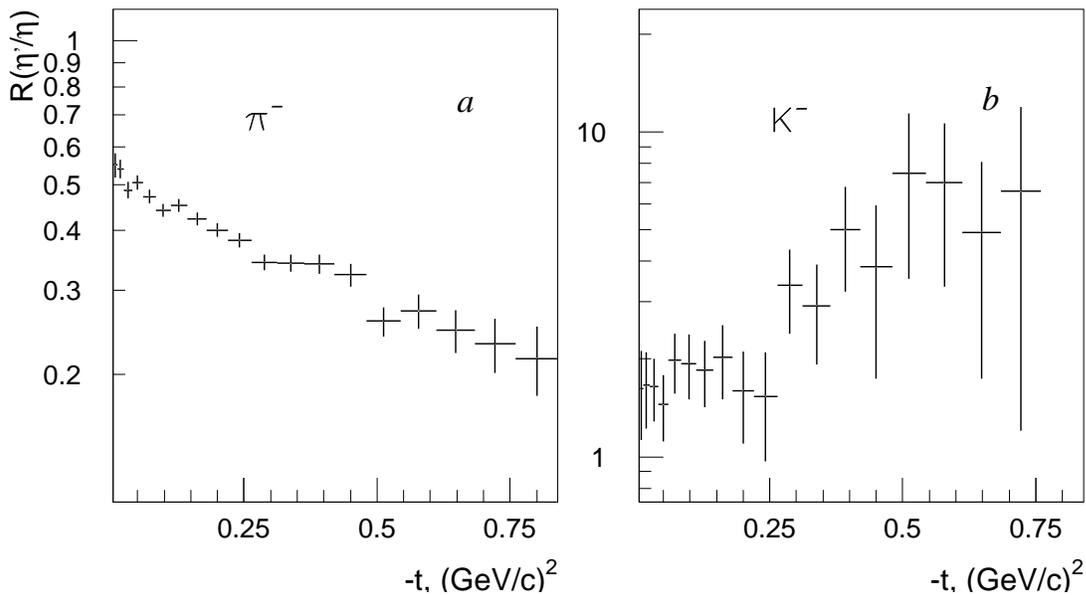} 
\caption{The differential cross section ratio  $R(\eta'/\eta)$ in 
the $\pi^{-}$ ({\it a}) and in the $K^{-}$  ({\it b} ) beams 
as a transfer momentum function.}
\label{fig2g_ratio}
\end{figure}

The  $R_{\pi}$ and $R_{K}$  as the function of $t$ are shown in Fig.~\ref{fig2g_ratio}.
These figures require an additional studies.
%
\section{Conclusion}

The ratio of the differential cross sections at $t' = 0$ (GeV/c)$^{2}$ in the $\pi^{-}$ beam equal
$R_{\pi}(\eta'/\eta) = 0.56 \pm 0.04$ and the mixing angle  is
$\phi_{P} = (36.8 \pm 1.0)^{o}$ in quark base.
For the $K^{-}$ beam the ratio $R_{K}(\eta'/\eta) = 1.30 \pm 0.16$
corresponds the mixing angle $\phi_{P} = (41.3 \pm 1.8)^{o}$ which is
incompatible with pion beam result at $\approx 2.5 \sigma$ confidence level.
We assume that the reason of this difference is the gluon content in $\eta'$ meson.
Two angle mixing scheme with gluon mixing angle $\psi_{G}$ result in gluon content 
$\sin^{2}{\psi_{G}} = 0.15 \pm 0.06$ for $\eta'$ meson.

\section{Acknowledgement}
We are grateful  professor A.K.~Likhoded  and A.V.~Luchinsky  for useful discussions.\\
This work was performed under RFBR Project 13-02-00898.
\end{document}